\documentstyle[10pt,a4]{article}
\newcommand{\bi}{\bibitem}
\begin{document}
\begin{center}
{\bf CONTINUUM LIMIT OF NONLINEAR DISCRETE SYSTEMS WITH LONG RANGE
INTERACTION POTENTIALS}
\end{center}
\vskip .6cm
\begin{center}
{\bf A.S. C\^ arstea\footnote {E-mail address ~~~ acarst@theor1.ifa.ro}, 
D. Grecu\footnote {E-mail address ~~~dgrecu@theor1.ifa.ro}, 
Anca Vi\c sinescu\footnote {E-mail address ~~~avisin@theor1.ifa.ro}}\\
Department of Theoretical Physics\\
National Institute of Physics and  Nuclear Engineering\\
"Horia Hulubei"\\ 
Bucharest, Romania\\
\end{center}
\vskip .6cm

\begin{abstract} 
 One-dimensional nonlinear lattices with harmonic long range
interaction potentials (LRIP) of inverse power type are studied.
For the nearest neighbour nonlinear interaction we shall consider
the anharmonic potential of the Fermi-Pasta-Ulam problem and the 
$\phi^3+\phi^4$ as well. 
The continuum limit is
obtained  following the method used by Ishimori [1], and several
Boussinesq and KdV-type equations are found with supplementary 
Hilbert transform terms. This nonlocal terms are introduced by the LRIP.
For the $\phi^3+\phi^4$ nearest neighbour potential, the continuum
approximation turns out to admit exact bilinearization in Hirota formalism.
Exact rational nonsingular solutions are found. The  
connection with perturbed KdV equations is also discussed, and
the dynamics of the KdV soliton is studied for one of these equations.

\end{abstract}
\vskip .6cm
\begin{center}
{\bf Introduction}
\end{center}
\vskip .5cm
Several years ago Ishimori [1] has discussed in the continuum approximation 
a one-dimensional nonlinear lattice with a long range interaction potential 
of Lennard-Jones type. This was done expanding the lattice displacements in 
Fourier series, performing the summations in the real space, and finally 
taking into account only the first terms in the wave vector expansion
- the long wave limit. His conclusion was that the long range
interaction manifests mostly in the linear dispersive terms, the
effect on the nonlinear terms being less dramatic. Several
integro-differential equations were obtained, some of them
containing an Hilbert transform.

From another point of view, in the context of wave propagation
in shallow water, Whitham [2] has extended the well known
Korteweg-de Vries equation by including a nonlocal (integral)
dispersive term. An example from plasma physics is given by the
nonlinear ion acoustic waves  where a Hilbert transform term is 
introduced into the KdV equation to describe the Landau damping [3].

Recently the interest in the study of nonlinear evolution
equations with non-local dispersive terms was stimulated by the
discovery that their solutions have very interesting picking and
breaking phenomena [4]. On the other hand it is well known that
several integro-differential equations of Benjamin-Ono (BO)
type [5] are completely integrable and the main feature of their
integrability character is the existence of rational solutions
(algebraic solitons).  

Following Ishimori [1] we shall consider a one-dimensional
lattice with long range harmonic interaction between the atoms,
the nonlinear interaction being restricted between nearest
neighbours. The long range harmonic interaction potential is written

$${1\over 2} \Sigma _{m,n}^{'} J_{mn} (u_{m} - u_{n})^2, \eqno(1)$$
where
$$ J_{mn}= J_{\vert m - n \vert}= {1\over 2} {J\over {\zeta (p)}}
{1\over {\vert m-n \vert ^{p}}}, \eqno (2)$$
and in the summation the term $ m=n $ is excluded. Here
$u_{n},u_{m}$ are the dimensionless lattice displacements at the
lattice points $m,n$,  $p \geq 2$ is an integer describing the spatial
decreasing of the long range interaction, and $ J $
measures its strength. In the definition of  $J_{mn}$ we
introduced the Riemann sum
$$\zeta (p) = \sum _{m=1}^{\infty} {1\over m^{p}}, \eqno(3)$$
and consequently
$$\Sigma _{m}^{'} J_{mn}=J. \eqno(4)$$

Also a long range potential with alternating sign will be
considered. For this $J_{mn}$ is given by
$$ J_{mn} = {1\over 2} {J\over \zeta (p)}{(-)^{\vert m-n+1 \vert} 
\over {\vert m-n \vert ^{p}}} \eqno(2a)$$
and in this case
$$\Sigma _{m}^{'} J_{mn}=J(1-{1\over 2^{p-1}}). \eqno(4a)$$

The nonlinear interaction is assumed only between nearest neighbours
$$\sum _{n} V(u_{n+1} - u_{n}) \eqno(5)$$
and the simplest form for $ V $ is 
$$ V(u) = {1\over 2} u^2 - {\alpha \over 3} u^3, \eqno(6a)$$
$ \alpha $ being a parameter describing the strength of the
anharmonic term. The expression (6) is the same as in the famous
Fermi-Ulam-Pasta problem. 
We shall discuss, also the following form
$$V(u)=\frac{1}{2}u^2+\frac{\alpha}{3}u^3+\frac{\beta}{4}u^4\eqno(6b)$$
which is an anharmonic potential introduced by M. Wadati [11]. Obviously
the continuum approximation for the equations of motion will involve 
Boussinesq type equations with long range interaction corrections.

Due to the inverse power form of the long range kernel, 
continuum limits of the 
long range interaction terms turn out to be in the form of a Hilbert
transform and its derivatives [1]. Also the continuum limits 
of the local terms, restricted at the waves in only
one direction, leads to the appearance of the KdV [7] equation for the
local potential (6a), and KdV+mKdV [11] equation for the local potential (6b).
In the first case, assuming $J$ to be small a perturbatively approach
is done in order to find what happened with the 1-soliton solution
of KdV type. It is found that for $p=2$ the first 5 conservation laws for KdV
1-soliton remain unchanged in the first order of perturbation.

In the second case the presence of the Hilbert transform
which is a nonlocal dispersive term leads us to the possibility
to consider rational solutions which appear in nonlinear integro-differential
equations such as BO-hierarchy [5] and nonlocal AKNS-hierarchy [5].
These rational solutions behave like solitons and interact elastically
with no phase shift. For {\it local} (1+1) dimensional nonlinear 
evolution equations both continuous and discrete
the rational solutions also exist but their character is rather singular
[12], [13], [14], [17]. They have no free parameters and there 
are very few cases
in which they are not singular [12], [15], [16] and, 
moreover, they interact
elastically with no phase shift [16]. 

It turns out that for $p=4$ and (6b)
the continuum limit of the equation of motion 
admits exact bilinearization in Hirota formalism
for certain values of the parameters $\alpha$ and $\beta$. 
The corresponding
rational solutions are exactly the same as for mKdV with 
nonvanishing
boundary conditions [12], [15]. 
In this way we are expecting that this
system to be completely integrable although it does not belong to
the BO-hierarchy or Nonlocal AKNS-hierarchy either.

\vskip .6cm
\begin{center}
{\bf Continuum approximation for the equations of motion}
\end{center}
\vskip .5cm
The equation of motion writes
$${d^2 u_{n} \over {dt^2}} = V^{'}(u_{n+1} - u_{n}) -
V^{'}(u_{n} - u_{n-1}) - 2 \Sigma _{m}^{'} J_{mn}(u_{n} -
u_{m}). \eqno(7)$$
In the linear approximation it becomes
$${d^2 u_{n} \over {dt^2}} = (u_{n+1} - 2 u_{n} + u_{n-1}) - 2 J
\left( u_{n} - {1\over {2 \zeta (p)}} \sum _{m=1} ^{\infty}
{1\over {m^{p}}} (u_{n+m} + u_{n-m})\right). \eqno(8)$$
Looking for solutions of the form
$$ u_{n} \simeq e^{i(kn-\omega t)} $$
the following dispersion relation results
$$\omega ^2 (k) = 4 \sin^2 {k\over 2} + 2 J {1\over {\zeta (p)}}
F_{p} (k), \eqno(9)$$
where
$$F_{p} (k) = \sum _{m=1} ^{\infty} {{1-\cos ~mk} \over m^{p}}. \eqno(10)$$
This is an even function of $k$ and can be calculated for each
value of $p$ [6]. In the following the results for several
values of $p$ will be presented.
\vskip .6cm
{\bf Case  p=2  }
\vskip .5cm

$$ F_{2}(k) = {\pi \over 2} \vert k \vert - {1\over 4} k^2,
~~~~~~\zeta (2) = {{\pi ^2}\over 6}. \eqno(11)$$
Then 
$$\omega ^2 (k) = 4~\sin^2~{k\over 2} + 2 J \left( {3\over \pi}
\vert k \vert - {3\over 2\pi ^2} k^2 \right), \eqno(12)$$
which in the long wave limit is approximated by
$$\omega^2 (k) \simeq {6J \over \pi} \vert k \vert + (1-{3J
\over \pi^2} )k^2 - \lambda k^4 + O(k^6), \eqno(13)$$
where $\lambda = {1\over 12} $ is a measure of the dispersion.
This dispersion relation corresponds to a linear partial differential
equation, namely
$$u_{tt}=a H(u_{x}) +c^2 u_{xx} +\lambda u_{xxxx}, \eqno(14)$$
where the subscript indicates the derivative with respect to the
corresponding variable, and the following notations were introduced
$$a={6J \over \pi},~~~~~~~~c^2 = 1- {3J \over {\pi^2}}. \eqno(15)$$
Here $H(f(x))$ is the Hilbert transform of the function $f(x)$
$$H(f(x)) = {1\over \pi} P \int_{-\infty}^{+\infty} {f(x^{'})
\over {x^{'}-x}} dx^{'}, \eqno(16)$$
and we have used the relation
$$H(e^{ikx}) = i~ sgn~ k~ e^{ikx}. \eqno(16a)$$

The contribution of the anharmonic term $\alpha u^3 $ in the
equation of motion is
$$-\alpha (u_{n+1} - u_{n} )^2 + \alpha (u_{n}-u_{n-1})^2 = -\alpha
(u_{n+1}-u_{n-1})(u_{n+1}-2 u_{n}+u_{n-1})$$
and in the continuum limit this goes to
$$ - 2 \alpha u_{x}u_{xx} = -\alpha (u_{x}^2)_{x}. \eqno(17)$$
Then the complete nonlinear partial differential equation in the
continuum limit is
$$ u_{tt}-c^2 u_{xx}+\alpha (u_{x}^{2})_{x}-\lambda u_{xxxx} -
aH(u_{x})=0. \eqno(18)$$
This is a Boussinesq type equation with a correction term in the
form of a Hilbert transform because of the long range harmonic
interaction.
\vskip .6cm
{\bf Case p=3} 
\vskip .5cm

In this case (see the Appendix)
$$F_{3}(k)=-{k^2\over 2} \ln ~\vert k \vert + {3 \over 4} k^2 - 
{1 \over 288} k^4 + O(k^6). \eqno(19)$$
The dispersion relation becomes
$$ \omega ^2 (k) = (1 + {3J\over 2\zeta (3)})k^2 - {J \over \zeta(3)} 
k^2 \ln~ \vert k \vert - {1\over 12 }\left( 1- {J\over {24 \zeta
(3)}}\right)
k^4 + O(k^6), \eqno(20)$$
and the corresponding Boussinesq type equation is
$$u_{tt} - c^2 u_{xx} + \alpha (u_{x}^2)_{x} - \lambda u_{xxxx} + a
T(u_{xxx}) = 0, \eqno(21)$$
where $$c^2 = 1 + {3J \over {2 \zeta (3)}}$$
$$\lambda = {1\over 12}(1- {J\over {24 \zeta(3)}})\eqno(22)$$
$$a= {J\over \zeta(3)} $$
and $T(f(x))$ is the following integral operator [1]
$$T(f(x))={1\over \pi} \int_{-\infty}^{+\infty} sgn~(x^{'} -
x)\left( \ln~\vert x^{'} - x \vert + \gamma \right) f(x^{'})
dx^{'} \eqno(23)$$
$\gamma $ being the Euler constant. In deriving the above equation we
have used 
$$T(e^{ikx}) = {\ln~ \vert k \vert \over {i k }} e^{ikx}. \eqno(23a)$$
\vskip .6cm
{\bf Case p=4}
\vskip .5cm
$$F_{4}(k)={\pi^2 \over 12} k^2 - {\pi \over 12} \vert k^3 \vert
+ {1\over 48} k^4, ~~~~~~~~\zeta (4)= {\pi^4 \over 90}, \eqno(24)$$
and the following equation is finally obtained.
$$u_{tt} - c^2 u_{xx} + \alpha (u^2_{x})_{x} - \lambda u_{xxxx}
- a H(u_{xxx})=0. \eqno(25)$$
Here $$c^2=1-{15J\over \pi^2},~~~~~~a={15J \over \pi^3},~~~~~~
\lambda ={1\over 12} (1- {45J \over \pi^4}). \eqno(25a)$$
\vskip .7cm

For an {\bf alternating in sign} long range interaction potential,
when $J_{mn}$ is given by (2a), the same technique gives
slightly different results. Instead of (10) $F_{p}(k)$ is now
defined as
$$F^{(a)}_{p}(k)=\sum^{\infty}_{m=1} (-1)^{m-1} {{1-
\cos~km}\over m^{p}}. \eqno(26)$$

As an example we shall consider the case $p=2$ when [6]
$$F^{(a)}_{p}(k) = {\pi^2 \over 4} k^2. \eqno(27)$$
Consequently the dispersion relation in the long wave limit
takes a very simple form
$$\omega ^2 (k) = c^2 k^2 - \lambda k^4 + O(k^6) \eqno(28)$$
$$c^2 = 1+ 3J$$
and the corresponding nonlinear evolution equation in this
continuum limit becomes
$$u_{tt}-c^2 u_{xx} + \lambda u_{xxxx}+ \alpha (u^2 _{x})_{x} = 0.\eqno(29)$$
No non-local corrections appear and the long range interaction
has a small influence modifying only the wave velocity.

It is well known since the paper of Zabusky and Kruskal [7] that
the KdV eq. is obtained as the uni-directional long time
approximation of the Boussinesq equation. We have to assume that
there is a competition between the dispersion, nonlinearity and
we have to add now the new term in the form of the Hilbert transform, 
competition which manifests at long times. In order to describe 
it we shall assume
$$\alpha \rightarrow \epsilon \alpha, ~~~~~\lambda \rightarrow \epsilon 
\lambda,~~~~a \rightarrow \epsilon a, \eqno(30)$$
where $\epsilon$ is a small parameter. The solution $u(x,t)$ is
written as

$$u(x,t) = f(\xi , T) +\epsilon u^{(1)}(x,t)\eqno(31)$$
where $$\xi = x-ct$$ $$T= \epsilon t \eqno(32)$$
are the "slow variables". Then in the first order in $\epsilon$
from (18) we get
$$u^{(1)}_{tt} - c^2 u^{(1)}_{xx} = 2cf_{_{\xi T}} - \alpha (f^2
_{_{\xi}})_{_{\xi}} + \lambda f_{_{\xi \xi \xi \xi }} + a H(f_{_{\xi}}).
\eqno(33)$$
As the right hand side depends only on $\xi $ and not on $\eta
=x+ct$ it has to vanish, in order to prevent the linear rising
in $\eta $ of $ u^{(1)}(x,t)$. Defining
$$q={1\over 3} f_{_{\xi }}$$
$$\tau = {T\over {2c}}, \eqno(34)$$
we finally obtain
$$q_{\tau}- 6 q q_{_{\xi}}+ \lambda q_{_{\xi \xi \xi }} + a H(q)=o \eqno(35)$$
which in a KdV equation with an additional term (long range
correction) in the form of a Hilbert transform.

If the same procedure is applied to the equation (25) we get
$$q_{\tau} - 6 q q_{_{\xi}} + \lambda q_{_{\xi \xi \xi}} - a H(q_{_{\xi
\xi}})=o. \eqno(36)$$
Here the correction term has the same form as the Hilbert
transform appearing in the Benjamin-Ono equation [5], which is
actually obtained if the dispersive term is dropped out.

A quite surprising result emerges in the case of long range
interaction with alternating signs. As shown before in this case
the continuum limit is the usual Boussinesq equation (29) and
from it the KdV eq. is immediately derived. It seems that the
alternating signs in the long range interaction potential (we
can call it an "antiferromagnetic" interaction) has a very small
effect on the nature of elementary excitations in the long wave
limit. This conjecture has to be verified for other values of
$p$ in (2a).
\vskip .6cm
\begin{center}
{\bf Perturbatively approach}
\end{center}
\vskip .5cm
Now for small values of the parameter "$a$" one can ask
ourselves in what way the soliton of the KdV eq. is perturbed.
We shall discuss briefly this problem for eq. (35) (we shall put
$\lambda =1 $) following one of the methods used to the
analytical description of soliton dynamics in nearly integrable
systems [8]. The approach is based on calculating the changes
( evolution in time ) of the first conservation laws. It is
easily seen from eq. (35) that the first two, the mass and
impulse, remain unchanged
$$\int_{- \infty}^{+ \infty } q(\xi, \tau ) d\xi = ct.$$
$$\int_{- \infty}^{+ \infty } q^{2}(\xi, \tau) d\xi = ct. \eqno(37)$$
In order to find the time evolution of the KdV soliton energy we
have to multiply eq. (35) by
$$3q^2 + q_{_{\xi}} {\partial \over {\partial \xi }}$$
and integrate over $\xi $. One obtains
$${dE_{k} \over  {d\tau }} + a \int_{- \infty }^{+ \infty } d\xi 
(3 q^2 + q_{_{\xi }} {\partial \over {\partial \xi }}) H(q(\xi))=0 \eqno(38)$$
where for the one soliton solution
$$q_{k}(\xi \tau)= - 2 k^2 sech^2 \left( k(\xi -4 k^2 \tau ) \right)$$
$$E_{k}= \int_{- \infty}^{+ \infty } \left( {1\over 2}
q^2_{_{\xi}}+q^3 \right) d\xi = - {32\over 5}k^5. \eqno(39)$$
The second term in (38) separates in two integrals, namely:
$$I_{1} = {3\over \pi} P \int_{-\infty}^{+\infty}
\int_{-\infty}^{+\infty} d\xi ~d\xi^{'}~{q^2(\xi)~q(\xi^{'}) \over
{\xi^{'} -\xi }} \eqno(40)$$
and
$$I_{2} = {1\over \pi} P \int_{-\infty}^{+\infty}
\int_{-\infty}^{+\infty} d\xi ~d\xi^{'}~{\partial q(\xi) 
\over \partial \xi}~ {\partial \over \partial \xi} \left( {q(\xi^{'}) 
\over {\xi^{'}-\xi}} \right). \eqno(41)$$
After a partial integration on $\xi $ in $I_{2}$ we get
$$I_{2} = - {1\over \pi} P \int_{-\infty}^{+\infty}
\int_{-\infty}^{+\infty} d\xi ~d\xi^{'}~{q_{_{\xi \xi}}(\xi)~q(\xi^{'}) 
\over {\xi^{'}-\xi}}. \eqno(41a)$$
But for the KdV soliton 
$$q_{_{\xi \xi}}=4k^2q +3q^2. \eqno(42)$$
Then using the fact that
$$\int_{-\infty}^{+\infty} \int_{-\infty}^{+\infty} d\xi d\xi^{'}
{q(\xi)~q(\xi^{'}) \over {\xi^{'} - \xi}} = 0, \eqno(43)$$
relation which was already used for proving the momentum
conservation (37), we obtain
$$I_{1} + I_{2} = 0 \eqno(44)$$
and consequently also the soliton energy is conserved in a first
order perturbation calculation.

We can perform the same procedure for the following conserved quantity:
$$T_{4}=5q^4+10qq_{\xi}^2+q_{\xi\xi}^2$$
Multiply eq (35) by:
$$20q^3+10q_{\xi}^2+20qq_{\xi}\frac{\partial}{\partial \xi}+2q_{\xi\xi}
\frac{\partial^2}{\partial {\xi}^2}$$
and integrating over $\xi$ using the same procedure it is found that
$T_{4}$ remains conserved. Lenghty but straightforward calculations
lead to the same result for the next conserved quantity:
$$T_{5}=21q^5+105q^2 q_{\xi}^2+21qq_{\xi\xi}^2+\frac{3}{2}q_{\xi\xi\xi}^2$$
We can proceed to any conserved quantity but the calculations are very much
involved. 

This is quite surprising because for an equation not very
different from (35), describing the Landau damping of an ion
acoustic wave [3], namely
$$n_{\tau} - 6 n n_{_{\xi}} + n_{_{\xi \xi \xi}} + a H(n_{_{\xi}}) = 0 
\eqno(45)$$
only the first conservation law (particle conservation) is not
destroyed, the other ones decaying in time.

In the same time we have to stress that our result is valid only
in the first order of a perturbation theory and the resulting
conclusions have to be understood only in this sense. More
elaborated techniques have to be used to go beyond this
approximation [8], [9].

That the situation is much more complicated was proven ten years
ago by Birnir [10]. Discussing an equation of the same form as
(35), with the coefficient $a$ in front of the Hilbert transform
a periodic function, satisfying mild conditions, he was able to
prove the existence of chaotic rational solutions. They look
like their integrable brothers, but are moving in a chaotic
fashion and also can disappear and reappear [10]. This result
shows how complex the problem of the perturbed KdV equation (35)
is.
\vskip .6cm
\begin{center}
{\bf Exact bilinearization}
\end{center}
\vskip .5cm
In this case the local potential is given by:
$$V(u)=\frac{1}{2}u^{2}+\frac{\alpha}{3}u^{3}+\frac{\beta}{4}u^{4}
\eqno(46)$$
where we assume $\beta\geq 0$
For $p=4$ the continuum approximation will have one more nonlinear term
due to the $u^{4}$ anharmonic term in the potential, so:
$$u_{tt}-c^{2}u_{xx}-\alpha(u_{x}^{2})_{x}-\beta(u_{x}^{3})_{x}-
\lambda u_{xxxx}-a Hu_{xxx}=0\eqno(47)$$

Following the same procedure, we consider:
$$u(x,t)=f(\xi,T)+\epsilon u^{(1)}(x,t)\eqno(48)$$
and one gets:
$$u^{(1)}_{tt}-c^{2}u^{(1)}_{xx}=2cf_{\xi T}+\alpha(f_{\xi}^{2})_{\xi}
+\beta(f_{\xi}^{3})_{\xi}+\lambda f_{\xi\xi\xi\xi}+a Hf_{\xi\xi\xi}=0
\eqno(48)$$
For $\tau=T/2c$ the equation becomes, after one integration:
$$f_{\tau}+\alpha f_{\xi}^{2}+\beta f_{\xi}^{3}+\lambda f_{\xi\xi\xi}+
a Hf_{\xi\xi}=0\eqno(49a)$$
which is a potential version of KdV+mKdV equation with nonlocal dispersion.
For the sake of simplicity we shall consider $J$ such that $\lambda=a=1$.
One can see for $\tau=T/2c$ and $q(\xi,\tau)=\frac{1}{3}f_{\xi}$, (48)
becomes:
$$q_{\tau}+6\alpha qq_{\xi}+27\beta q^2 q_{\xi}+q_{\xi\xi\xi}+Hq_{\xi\xi\xi}
=0\eqno(49b)$$

In order to bilinearize the equation (49a) we shall use the following
substitution:
$$f(\xi,\tau)=i\log{\frac{g_{+}(\xi,\tau)}{g_{-}(\xi,\tau)}}\eqno(50)$$
where
$$g_{-}(\xi,\tau)=\prod_{n=1}^{N}(\xi-    z_{n}(\tau))\eqno(51a)$$, 
$$g_{+}(\xi,\tau)=\prod_{n=1}^{N}(\xi-z_{n}^{*}(\tau))\eqno(51b)$$
and $Im z_{n}(\tau)\geq 0$

With this ansatz

$$Hf_{\xi\xi}=-\frac{\partial^{2}}{\partial \xi^{2}}\log{(g_{+}g_{-})}$$
and introducing in (49a) the following expression appears:
$$
g_{+}^{3}g_{-}^{3}[(i{\bf D}_{\tau}-{\bf D}_{\xi}^{2}+i{\bf D}
_{\xi}^{3})g_{+}\bullet g_{-}]
-(\alpha-1)g_{+}^{2}g_{-}^{2}({\bf D}_{\xi}g_{+}\bullet g_{-})^{2}-$$
$$-3ig_{+}^{2}g_{-}^{2}({\bf D}_{\xi}g_{+}\bullet g_{-})
({\bf D}_{\xi}^{2}g_{+}\circ
g_{-})+i(2-\beta)g_{+}g_{-}
({\bf D}_{\xi}g_{+}\bullet g_{-})^{3}=0\eqno(52)$$ 

where ${\bf D}_{x}^{n}f\bullet g=
(\partial_{x}-\partial_{y})^{n}f(x)g(y)\mid_{x=y}$
are the Hirota bilinear operators.

If $\alpha=1$, $\beta=2$ and decouples (52) one gets:
$$(i{\bf D}_{\tau}-{\bf D}_{\xi}^{2}+i{\bf D}_{\xi}^{3})g_{+}\bullet g_{-}=0$$
$${\bf D}_{\xi}^{2}g_{+}\bullet g_{-}=0$$
which relaxes to

$$({\bf D}_{\tau}+{\bf D}_{\xi}^{3})g_{+}\bullet g_{-}=0$$
$${\bf D}_{\xi}^{2}g_{+}\bullet g_{-}=0$$
and it is nothing but bilinear form for mKdV equation. There is no solution
of the form (51a), (51b).

If $\beta=2$ and $\alpha\not=1$ one obtains:
$$(i{\bf D}_{\tau}-{\bf D}_{\xi}^{2}+i{\bf D}_{\xi}^{3})g_{+}\bullet g_{-}=0$$
$$\left({\bf D}_{\xi}^{2}-\frac{i}{3}(\alpha-1){\bf D}_{\xi}\right)
g_{+}\bullet g_{-}=0\eqno(53)$$

The equation (53) is equivalent with
$$\left(i{\bf D}_{\tau}-\frac{i}{3}(\alpha-1){\bf D}_{\xi}+
i{\bf D}_{\xi}^{3}\right)g_{+}\bullet g_{-}=0$$
$$\left({\bf D}_{\xi}^{2}-\frac{i}{3}(\alpha-1){\bf D}_{\xi}\right)
g_{+}\bullet g_{-}=0\eqno(53)$$

which represents the bilinear form for the following mKdV equation
with nonvanishing boundary condition:
$$u_{\tau}+3(\alpha-1)u^{2}u_{\xi}+u_{\xi\xi\xi}=0\eqno(54)$$
and $u\rightarrow -1/3$ when $\xi\rightarrow \pm\infty$. 
This equation is completely integrable and it admits $N$-soliton solution.
By the long-wave limit procedure of Ablowitz and Satsuma it admits also
nonsingular real rational solutions in the form (51a) and (51b)
and accordingly they are solutions of (49). Due to the fact that
(54) admits rational nonsingular soluitons for every odd N, we
can conclude that the set of rational solutions for (49) is the same
with the set of rational solutions for (54).
The 1-rational and 3-rational solutions for (49a) are given by:

$$g_{\pm}(\xi,\tau)=\theta(\xi,\tau)\mp\frac{3i}{\alpha-1}+d$$
and
$$g_{\pm}(\xi,\tau)=\theta(\xi,\tau)^3+12\tau-\frac{27}{(\alpha-1)^2}
\theta(\xi,\tau)\mp\frac{9i}{\alpha-1}
\left[\theta(\xi,\tau)^2+\frac{9}{(\alpha-1)^2}\right]+d$$
where $\theta(\xi,\tau)=(\xi+(\alpha-1)\tau/3+c)$ and 
$c$, $d$ are arbitrary constants. 
Thus, $q=i\partial_{\xi}\log(g_{+}/g_{-})$ is a nonsingular rational solution
We can see that only for $N=1$
the solution is a solitary wave. The velocity is fixed 
and the 3-rational solution is not a superposition of solitary waves.

\vskip .6cm
{\bf Appendix}
\vskip .5cm

In calculating $F_{k}$ we start from
$$F_{3}(k) = \int_{0}^{\infty} \left( \sum_{m=1}^{\infty} 
{\sin (mk^{'}) \over m^2} \right) dk^{'} $$
and the integral
$$  \sum_{m=1}^{\infty} {\sin (mk) \over m^2} = - \int_{0}^{k} \ln
(2 \sin {t\over 2}) dt. $$
We have
$$F_{3}(k) = - \int_{0}^{k} dk^{'} \int_{0}^{k^{'}} 
\ln (2 \sin {t\over 2}) dt $$ 
$$ = - \int_{0}^{k} (k - t) \ln (2 \sin {t\over 2}) dt $$ 
$$ = - {k^2 \over 2} \ln ( 2 \sin {k\over 2})
+ {1\over 2} \int_{0}^{k} (kt - {t^2\over 2}) \cot {t\over 2} dt.$$
But
$$\ln (2 \sin {k\over 2}) \simeq \ln k + \ln ( 1 - {k^2\over 24} + ...)
\simeq \ln k - {k^2\over 24} + O(k^4) $$
and $$ \cot {t\over 2} \simeq {2\over t} (1 - {t^2\over 12} + O(t^4)).$$
Then, by straightforward integration, we get $(k > 0)$
$$F_{3}(k) = - {k^2\over 2} \ln k + {3\over 4} k^2 +
{1\over 288} k^4  +O(k^5).$$
%\vskip .5cm

\end{document}